\documentclass[twocolumn,english,prb,superscriptaddress,showpacs, reprint]{revtex4}
\usepackage[T1]{fontenc}
\usepackage[latin9]{inputenc}
\usepackage{amstext}
\usepackage{graphicx}
\usepackage{amssymb}

\makeatletter

\newcommand{\noun}[1]{\textsc{#1}}
\providecommand{\tabularnewline}{\\}

\@ifundefined{textcolor}{}
{%
 \definecolor{BLACK}{gray}{0}
 \definecolor{WHITE}{gray}{1}
 \definecolor{RED}{rgb}{1,0,0}
 \definecolor{GREEN}{rgb}{0,1,0}
 \definecolor{BLUE}{rgb}{0,0,1}
 \definecolor{CYAN}{cmyk}{1,0,0,0}
 \definecolor{MAGENTA}{cmyk}{0,1,0,0}
 \definecolor{YELLOW}{cmyk}{0,0,1,0}
 }

\makeatother

\usepackage{babel}

\begin{document}

\title{Designed Metamagnetism in CoMnGe$_{1-x}$P$_{x}$ }

\author{Z. Gercsi}

\affiliation{Dept. of Physics, Blackett Laboratory, Imperial College London, London
SW7 2AZ UK}

\author{K. Hono}

\affiliation{Magnetic Materials Center, National Institute for Materials Science
(NIMS), 1-2-1 Sengen, Tsukuba 305-0047, Japan}

\author{K.G. Sandeman}

\affiliation{Dept. of Physics, Blackett Laboratory, Imperial College London, London
SW7 2AZ UK}

\pacs{75.30.Sg,75.30.Kz, 75.80.+q, 75.30.Et}
\begin{abstract}
We extend our previous theoretical study of Mn-based orthorhombic
metamagnets to those that possess large nearest neighbour Mn-Mn separations
($d{}_{1}\gtrsim3.22$~Å). Based on our calculations, we design and
synthesize a series of alloys, CoMnGe$_{1-x}$P$_{x}$, to experimentally
demonstrate the validity of the model. Unusually, we predict and prepare
several metamagnets from two ferromagnetic end-members, thus demonstrating
a new example of how to vary crystal structure, within the \emph{Pnma}
symmetry group, to provide highly tunable metamagnetism. 
\end{abstract}
\maketitle

\section{Introduction}

Manganese-based orthorhombic (\emph{Pnma}) binary and ternary alloys
are of fundamental research interest as they often exhibit complex,
non-collinear magnetic structures that can be tuned by temperature,
pressure and applied magnetic field. Some well-known examples are
the set of fan, helical and cycloidal spin structures found in the
(\emph{H,T}) phase diagram of MnP \citep{nagamiya1967,Kallel1974,Niziol1982},
the cycloidal antiferromagnetism (AFM) of IrMnSi \citep{ERIKSSON2005}
and helical metamagnetism of CoMnSi \citep{Nizol1978,Niziol1982}.
Various theoretical explanations have previously been suggested to
describe the mechanisms responsible for the noncollinear magnetism
of such materials. Some refer to competing symmetric and asymmetric
exchange interactions \citep{Kallel1974,Dobrzynski1989}; others to
conduction-mediated indirect (RKKY) exchange \citep{ElliottRKKY}.
Another potential cause put forward is band crossing and appropriate
Fermi surface topology (nesting) \citep{ERIKSSON2005,Eriks_bandcross}. 

One of the most feature-rich materials of this kind is CoMnSi on account
of its pronounced magnetic field-induced tricritical metamagnetism
and associated negative magnetocaloric effect (MCE)\citep{SandemanPRB}.
Our recent high-resolution neutron diffraction (HRPD) study \citep{AlexHRPD}
uncovered a giant magneto-elastic coupling within the antiferromagnetic
ground state of this system. It occurs as a change of up to 2\% in
nearest-neighbour Mn-Mn separations $d{}_{1}$, $d{}_{2}$ on heating.
The field-induced tricriticality of this system can thus be understood
as the result of tuning the metamagnetic critical temperature with
an applied magnetic field to the point at which it coincides with
this native giant magneto-elasticity. 

Using Density Functional Theory (DFT) we recently examined the importance
of the same Mn-Mn separations in determining the occurrence of different
magnetic groundstates across several Mn-based orthorhombic (\emph{Pnma})
systems. \citep{Gercsi_MnP}. By applying hydrostatic expansion and
compression to a prototype model MnP alloy, we found a stability criterion
for the appearance of an AFM groundstate, rather than the usual FM
state seen in MnP. This direct relation between Mn-Mn separation and
magnetic groundstate can explain the energetic proximity of FM and
AFM states in materials such as CoMnSi \citep{Nizol1978,Niziol1982},
MnAs$_{\text{1-x}}$P$_{\text{x}}$ \citep{Fjellvag198429}, (Fe$_{\text{1-x}}$Co$_{\text{x}}$)MnP
\citep{CoFeMnP1} and NiMnGe$_{\text{1-x}}$Si$_{\text{x}}$\citep{Bazela1981}
where the nearest-neighbour Mn-Mn distances are close to a critical
separation of 2.95~Å$\lesssim d{}_{1}\lesssim$3.05~Å. 

Although the latter model was computed in a large interval of 2.5~Å$\lesssim d{}_{1}\lesssim$3.22~Å
to cover many of the relevant compositions in the literature, it may
lead to a misinterpretation of a technologically relevant alloy with
a larger $d{}_{1}=3.4$~Å spacing: CoMnGe. CoMnGe is a collinear
ferromagnet with a tendency to form a metastable hexagonal structure
upon rapid cooling\citep{Johnson75}. In this Article, we extend our
previous theoretical analysis \citep{Gercsi_MnP} towards larger Mn-Mn
separations to explain the re-appearance of ferromagnetism in alloys
with $d{}_{1}\gtrsim$3.37~Å. The importance of the correct theoretical
description of latter composition is due its large magnetocaloric
effect around room temperature\citep{Hamer2009,MnCrCoGe,CoMnGeB}. 

In this Article we first show the striking re-appearance of a FM groundstate
at large interatomic Mn separations in Mn-based \emph{Pnma} alloys
where $d{}_{1}\gtrsim$3.37Å, thereby accounting for the magnetic
properties of CoMnGe. Secondly, and significantly, we have designed
a new alloy series, CoMn(P,Ge) based on our extended model in order
to test and demonstrate its validity and in particular the dominance
of the ($d{}_{1}$) Mn-Mn separation in determining the magnetic groundstate
of the series of Mn-containing \emph{Pnma }alloys. We show that metamagnetism
can be derived, unusually, by inter-doping two ferromagnetic end-compositions
in order to bring $d{}_{1}$ to the critical regime where antiferromagnetism
and ferromagnetism are similar in energy. 

The remainder of the Article is organised as follows: first in Sec.
\ref{sec:Theoretical-Results}, the theoretical results calculated
by applying DFT to the prototype MnP alloy are given. Based on this
model, we present the structural and metamagnetic properties of purposefully
designed CoMnGe$_{1-x}$P$_{x}$ alloys in Sec. \ref{sec:Experimental-Results}.
Finally, a summary is made and conclusions are drawn in Sec. \ref{sec:Conclusions}.

\section{Theoretical \label{sec:Theoretical-Results}}

In our previous work we considered what we term the {}``prototype''
binary MnP (\emph{Pnma}) alloy and calculated the effect of isotropic
lattice expansion and compression on hypothetical non-magnetic (NM),
ferromagnetic (FM) and antiferromagnetic (AFM) states by using the
general gradient aproximation method (GGA-DFT) implemented in VASP
\citet{VASP}. We found the critical lattice parameters where a crossover
from one magnetic state to another can occur \citep{Gercsi_MnP}.
In that study, a single unit cell consisting of 8 atoms (4 Mn and
4 P) was used, which allowed three different collinear antiferromagnetic
configurations (AFM1, AFM2 and AFM3) and a collinear ferromagnetic
(FM) one to be constructed. In the interval of 2.5~Å$\lesssim d{}_{1}\lesssim$3.22~Å
we were able to predict a transition in the zero temperature magnetic
structure from NM to FM, and finally to AFM as a function of expanding
lattice parameters. A detailed description of the DFT calculations
is given in that work \citep{Gercsi_MnP}. Here we extend our simple
binary model to $d{}_{1}$>3.22~Å values by further hydrostatic expansion
in order to interpret ferromagnetism in CoMnGe where $d{}_{1}=3.4$~Å
in the current model. 

\begin{figure}
\includegraphics[width=8.5cm]{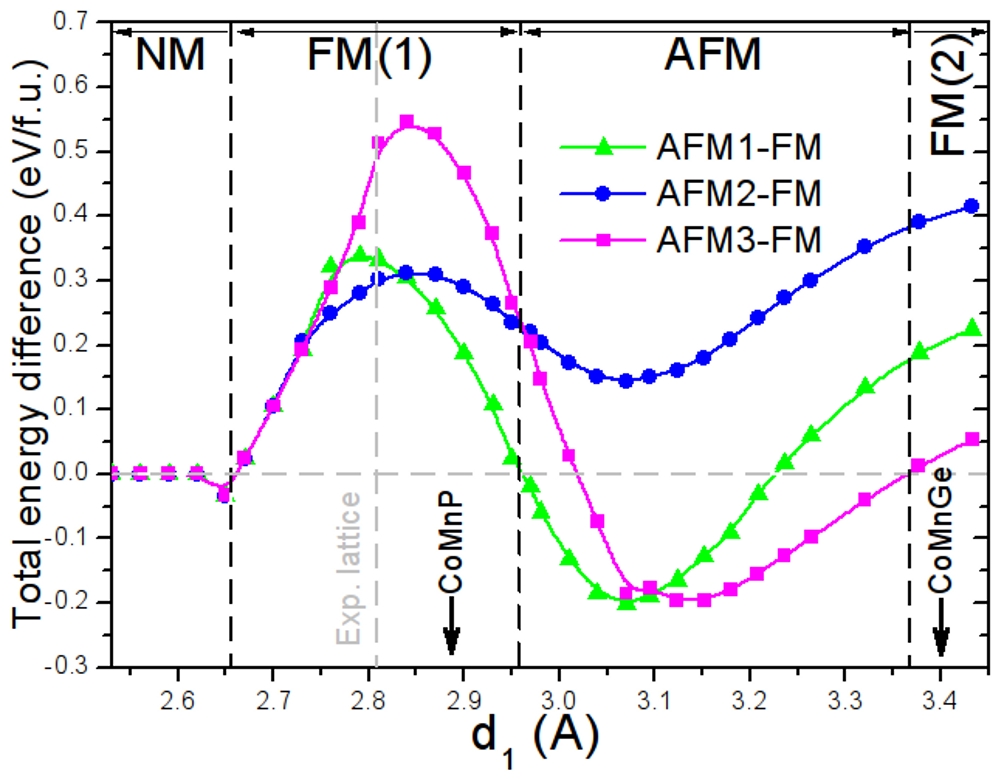}

\caption{\label{fig:StabilityandM} (Color online) Stability of possible collinear
magnetic structures, relative to ferromagnetism, within a single unit
cell of MnP as a function of $d{}_{1}$ Mn-Mn separation. AFM configurations
become stable where $\triangle E_{AFM-FM}<0$. The vertical dashed
line represents the experimental (strain-free, $\varepsilon=0$) lattice
of MnP. We see that the FM state is first destabilised by lattice
expansion, and then becomes stable again at large $d{}_{1}$values:
FM(2).}

\end{figure}

Fig.~\ref{fig:StabilityandM} shows the difference in energy between
the possible collinear AFM and FM magnetic states $(\triangle E_{Tot}=E_{AFM}-E_{FM})$
as a function of Mn-Mn separation. Using this comparison scheme a
non-FM state becomes most favourable when it has the most negative
value of $\triangle E_{Tot}$. On the left hand side of Fig. \ref{fig:StabilityandM},
the large compression causes a strong overlap of\emph{ d}-orbitals,
and the broad \emph{d}-\emph{d} hybrid bands thus formed cannot support
spontaneous magnetisation. In agreement with experimental findings,
the FM state is the groundstate of MnP and is stable for intermediate
deformations of the lattice. On further expansion of the lattice parameters,
first the AFM1 type magnetic structure (at around $d{}_{1}$$\thicksim$2.97~Å)
and then the AFM3-type ordering (at $d{}_{1}$$\thicksim$3.1~Å)
become energetically favorable. However the extended study presented
here shows that AFM3-type ordering ceases to be the most stable magnetic
state for large lattice expansion, and eventually the collinear FM
state is once again the groundstate for $d{}_{1}$$\gtrsim$3.37~Å.

In order to experimentally prove the validity of the model, we carefully
selected two collinear FM compositions with lattice parameters from
the different FM regions of the stability plot in Figure~\ref{fig:StabilityandM}.
Our choices were CoMnP from the FM(1) regime ($d{}_{1}$$\thicksim$2.95Å\citep{FruchCoMnP})
and CoMnGe ($d{}_{1}$$\thicksim$3.4Å\citep{Niziol1982} ) from the
larger Mn-Mn separation (FM2) zone. Our hypothesis is that progressive
substitution of one p element for another in CoMnGe$_{1-x}$P$_{x}$,
without changing the 3d-element concentration, can cause the appearance
of metamagnetism in particular compositions of the series. This replacement
of large germanium atoms by the much smaller phosphor atoms should
result in a shrinkage of lattice. From Figure~\ref{fig:StabilityandM}
we expect that the decreasing $d{}_{1}$ separation will lead to the
destabilization of FM state in competition with the AFM one at a certain
P/Ge ratio. It should then be possible to manipulate the magnetic
state of the energetically metastable alloys by changing temperature
or applied magnetic field. 

In order to identify the key factors that can lead to the substantial
changes in magnetic groundstate, we first calculate the electronic
band structure and magnetic moment of CoMnGe$_{1-x}$P$_{x}$ alloys
with x=0, 0.25, 0.5, 0.75 and 1 in a collinear FM state. The calculated
magnetic moments are given in Table \ref{tab:Mteo}. The magnetic
moments on Mn (2.68$\mu_{B}$) and Co (0.28$\mu_{B}$) sites in the
CoMnP alloy agrees well with previous calculations based on the KKR
method with coherent potential approximation (CPA) by Zach and co-workers~\citep{CoFeMnP1_Dos}.
Furthermore, the partial replacement of P by Ge results in a progressive
increase of magnetic moment on both 3d elements, leading to an increased
$M{}_{Tot}$ of up to 3.58~$\mu_{B}$ for CoMnGe. A small negative
moment induced on the p-block elements is also observed. 

\begin{figure}
\includegraphics[width=85mm]{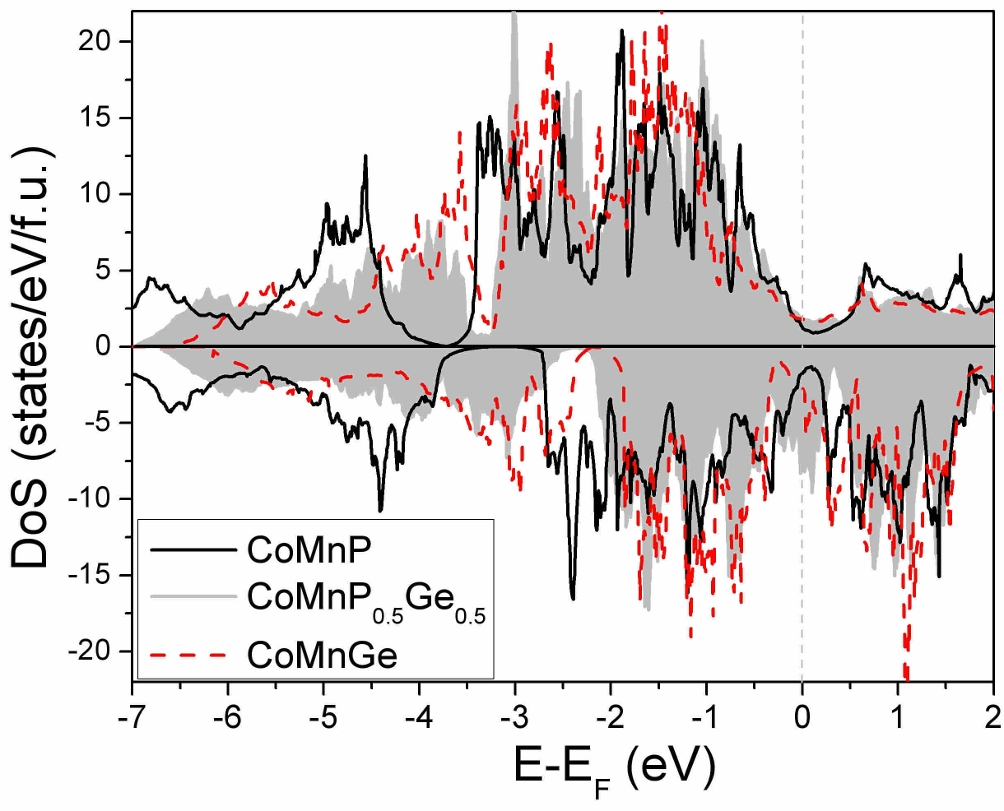}

\caption{\label{fig:DOS} (Color online) Collinear FM total density of states
for CoMnGe$_{1-x}$P$_{x}$ with x=0, 0.5 and 1. The Fermi energy
falls into the hybridization gap for both CoMnP and CoMnGe, but shifted
for CoMnP$_{0.5}$Ge$_{0.5}$ resulting in a large $N_{Tot}(E_{\mathbb{F}})$. }

\end{figure}

The FM total density of electronic states (DoS) of the two end compositions,
CoMnP and CoMnGe, are plotted, together with CoMnP$_{0.5}$Ge$_{0.5}$
in Fig.~\ref{fig:DOS}. Although the value of total density of states
at the Fermi level, $N_{Tot}(E_{\mathbb{F}})=N_{\Downarrow}(E_{\mathbb{{\normalcolor F}}})+N_{\Uparrow}(E_{\mathbb{{\normalcolor F}}})$
exhibits a large change with composition, each total DoS possesses
the same features over a large extent of energy range. The main difference
is the location of these features, and in particular the location
of a pseudogap-like feature in the DoS. Using CoMnP as reference,
if the energy scale of the minority DoS is shifted by about +0.25
eV for CoMnGe and about -0.3 eV for CoMnP$_{0.5}$Ge$_{0.5}$, not
only would the pseudogap fall at $E_{\mathbb{F}}$ but most of the
exchange-split DoS peaks of Mn and Co would line up at around the
same position. The large $N_{Tot}(E_{\mathbb{F}})$ in CoMnP$_{0.5}$Ge$_{0.5}$
(Table \ref{tab:Mteo}) suggests the instability of the collinear
FM state in this composition. A possible scenario that stabilizes
the noncollinear state through the formation of hybridization gap
at the Fermi energy is described by \citet{Eriks_bandcross}. We recently
showed the relevance of this mechanism in a noncollinear DFT study
on the metamagnet CoMnSi \citep{AlexHRPD}.

In the following sections, we are going to demonstrate the validity
of our theoretical prediction of metamagnetism in these Mn-based \emph{Pnma}
alloys through magnetic and structural results on an experimentally
synthesized CoMnGe$_{1-x}$P$_{x}$ series of alloys.

\begin{table}
\begin{tabular}{|c|c|c|c|c|c|}
\hline 
x & M$_{Mn}$ & M$_{Co}$ & M$_{Si/Ge}$ & M$_{Total}$ & $N_{Tot}(E_{\mathbb{F}})$\tabularnewline
\hline
\hline 
0 & 3.05 & 0.6 & -0.07 & 3.58 & 5.0\tabularnewline
\hline 
0.25 & 2.83 & 0.45 & -0.06 & 3.22 & 10.0\tabularnewline
\hline 
0.5 & 2.89 & 0.47 & -0.07 & 3.29 & 9.5\tabularnewline
\hline 
0.75 & 2.71 & 0.33 & -0.07 & 2.97 & 7.5\tabularnewline
\hline 
1 & 2.68 & 0.28 & -0.07 & 2.89 & 1.7\tabularnewline
\hline
\end{tabular}

\caption{\label{tab:Mteo} Calculated magnetic moments ($\mu_{B}$) and $N_{Tot}(E_{\mathbb{F}})$
(states/eV/f.u.) for CoMnGe$_{1-x}$P$_{x}$.}
 
\end{table}

\section{Experimental \label{sec:Experimental-Results}}

\subsection{Experimental details}

Samples of CoMnGe$_{1-x}$P$_{x}$ with $x=0.25,0.4,0.5,0.55,0.6$
and $0.75$ were prepared in a quartz nozzle by an induction melting
technique, using Co$_{2}$P (99.9\%) and Mn$_{3}$P$_{2}$(99.9\%)
master alloys mixed together with high purity Co (99.97\%), Mn (99.99\%)
and Ge (99.9999\%) elements in the required proportions. The alloys
were cast into a copper mold under protective Ar atmosphere. The ingots
were then sealed in quartz tube under protective He atmosphere and
a homogenization at 1000$^{0}$C for 24 hours and annealing treatment
at 800$^{0}$C for 72 hours then followed. The samples thus obtained
were crushed into fine powder in order to determine their crystal
structure using X-ray diffraction (XRD) with Cu $K\alpha$ radiation.
Structural (Rietveld) refinement of the data was carried out the using
the\noun{ Fullprof}~\citep{Fullprof} program. A microstructural
and compositional analysis was carried out using a Carl Zeiss 1540EsB
scanning electron microscope (SEM). Magnetic properties of the samples
were studied in a Quantum Design MPMS system.

\subsection{Crystal Structure \label{sub:Crystal-Structure} }

\begin{figure}
\includegraphics[width=85mm]{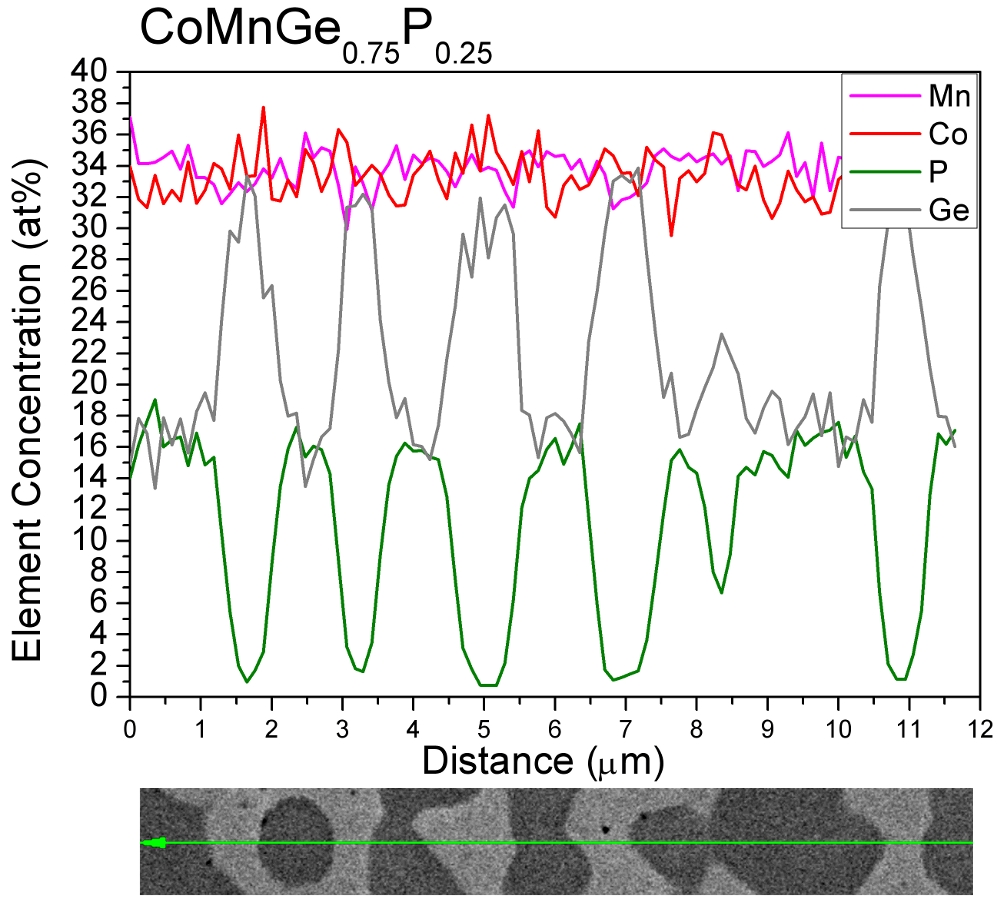}\caption{\label{fig:Sem1} Representative SEM micrograph and corresponding
elemental mapping from EDX of CoMnGe$_{0.75}$P$_{0.25}$. }

\end{figure}
Both CoMnGe and CoMnP alloys crystallize in the orthorhombic (\emph{Pnma},
62) structure in which the elements occupy general 4c ($x,\frac{1}{4},z$)
crystallographic positions. XRD analysis of the CoMn(Ge,P) samples
revealed the formation of this orthorhombic structure in all compositions.
Furthermore, in the samples with $x=0.25,0.4,0.5$ and $0.75$, extra
reflections in the diffraction pattern also appear that can be ascribed
to the hexagonal Ni$_{\text{2}}$$ $In-type (\emph{P}6$_{\text{3}}$/mmc,
194) lattice structure. The appearance of the higher symmetry hexagonal
phase is often observed in similar alloy systems because the orthorhombic
structure can be regarded as a distortion from this hexagonal structure
and the two structures can be interrelated as follows: $b_{ortho}=a_{hex}$
and $c_{ortho}=\sqrt{3}\times a_{hex}$. 

The importance of this latter correlation has been exploited in several
Mn-based \emph{Pnma} systems. In the pseudo-binary Mn$_{\text{1-x}}$Fe$_{\text{x}}$As
alloys the sharp, first order type magnetostructural transformation
(orthorhombic$\Longleftrightarrow$hexagonal)\citep{Fjellvag_MnFeAs}
is also accompanied by a {}``colossal'' MCE \citep{MnFePNature2006}.
The magnitude of the useful magnetic entropy change is, however, now
strongly contested~\citep{Balli_contraCampos}.A similar magnetostructural
transition in CoMnGe-based ternary compositions was reported by Kanomata
and co-workers who observed a large, $\thicksim$5.3\%, volume change
\citep{KanomataCoMnGe1,KanomataCoMnGe2}. Theoretical calculations
also revealed that the Co vacancy-induced phase transformation is
due to a high moment to low moment magnetic transition accompanied
by a large magnetovolume effect originating from the change of the
coupling distance between the principal magnetic atoms \citep{Wang_CoMnGe}.
As an extension of this study, both Hamer~\citep{Hamer2009}and Trung~\citep{CoMnGeB,MnCrCoGe}
recently demonstrated that the transitions can be fine-tuned by pseudo-ternary
additives (Sn, B or Cr elements) around room temperature for an enhanced
magnetocaloric effect. The appearance of the hexagonal structure at
room temperature in our samples can therefore be understood as a first
order transformation from the low temperature orthorhombic structure
to a high temperature hexagonal one.

\begin{figure}
\includegraphics[width=85mm]{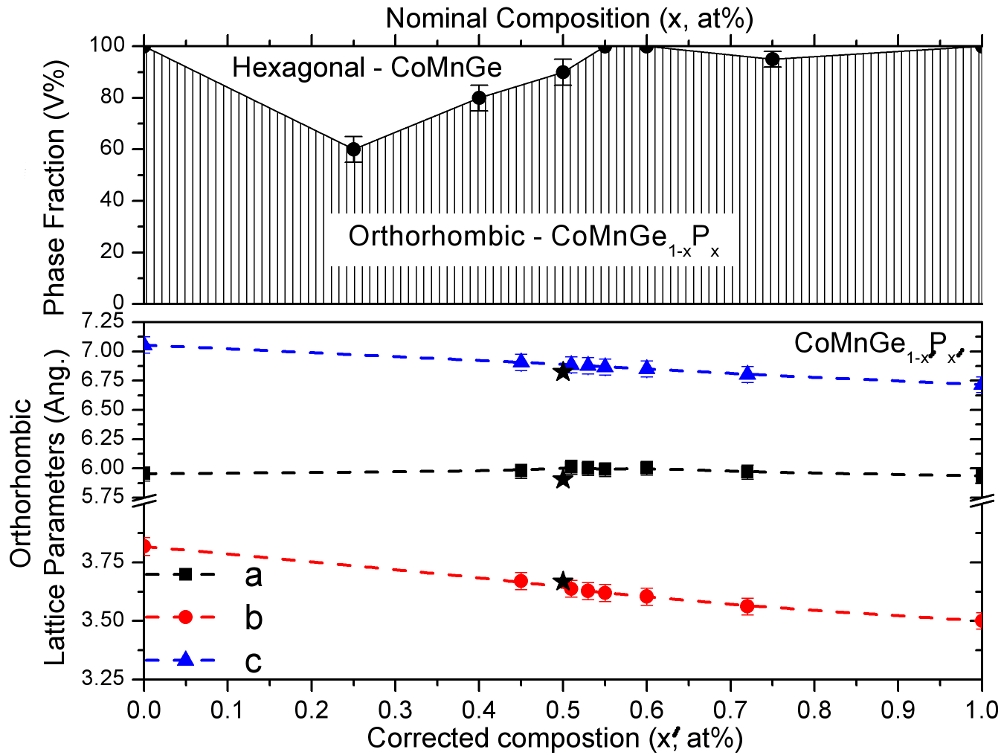}\caption{\label{fig:XRD_abc} (Color online) Volume fraction of the orthorhombic
CoMnGe$_{1-x}$P$_{x}$ and hexagonal CoMnGe phases as a function
of nominal composition (top) and lattice parameters of the main orthorhombic
phase as a function of corrected phosphor concentration (bottom).
The lattice parameters of CoMnSi, a metamagnet, are added (star symbols)
for comparison. }

\end{figure}

In order to clarify the composition of the hexagonal phase, a representative
secondary electron (SE) SEM micrograph was taken from the CoMnGe$_{0.75}$P$_{0.25}$
alloy and is shown in Fig.~\ref{fig:Sem1}a. The two different structural
regions are identified in accordance with the XRD results. EDX elemental
mapping revealed that the two phases have substantially different
atomic compositions. Our investigations show that the orthorhombic
phase is poor in Ge (and consequently enriched in P) whereas the second
phase is enriched in Ge and is therefore close in composition to stoichiometric
CoMnGe. An atomic composition profile taken along the direction of
the arrow indicated in Fig.~\ref{fig:Sem1}b shows quantitatively
the compositional difference between the two structures. These findings
are direct evidence for a compositional phase separation of the quaternary
alloy rather than for a second-order type transformation of the single
quaternary composition with temperature. 

The results of the quantitative Rietveld analysis in Fig. \ref{fig:XRD_abc}
(top) reveal the formation of single phase orthorhombic structures
in the alloys rich in P (\emph{x}>0.5). Fig. \ref{fig:XRD_abc} (bottom)
summarizes the lattice parameter of the orthorhombic structure as
a function P content. (The composition values ($x{}^{\prime}$) given
in the lower figure are corrected based on the quantitative analysis.)
The \emph{b} and \emph{c} lattice constants show decrease continuously
with Ge addition while the \emph{a} parameter stays nearly constant
and until the lattice parameters with \emph{x}$\thicksim$0.5 become
close to those of the metamagnet CoMnSi. 

In the next section, we will demonstrate the occurrence of metamagnetism
in these Mn-based alloys that have appropriately designed lattice
parameters.

\subsection{Magnetic properties }

As we demonstrated in Sec. \ref{sub:Crystal-Structure}, the lattice
parameters of the CoMnP$_{1-x}$Ge$_{x}$ alloy can be tuned towards
those of the CoMnSi metamagnet. In the present section, we will show
that this structural engineering allows us to prepare metamagnetic
quaternaries, even though the end alloys (CoMnP and CoMnGe) are ordinary
ferromagnets. Based on the \emph{a}, \emph{b} and \emph{c} lattice
parameters of the CoMnGe$_{1-x}$P$_{x}$ alloys, one would expect
progressive change from collinear ferromagnetism to non-collinear
anti-ferromagnetism with the replacement of P by Ge in the system.
At around \emph{x}$\thicksim$0.5 the structure should resemble that
of the CoMnSi metamagnet and once the composition is on the P-rich
end of the series, ferromagnetism should re-appear. Indeed, we have
found this scenario to be fulfilled; however the structural phase
separation into a ternary CoMnGe and a residual CoMn(Ge,P) phase in
compositions that are rich in Ge hinders a simple interpretation.%
\begin{figure}
\includegraphics[width=85mm]{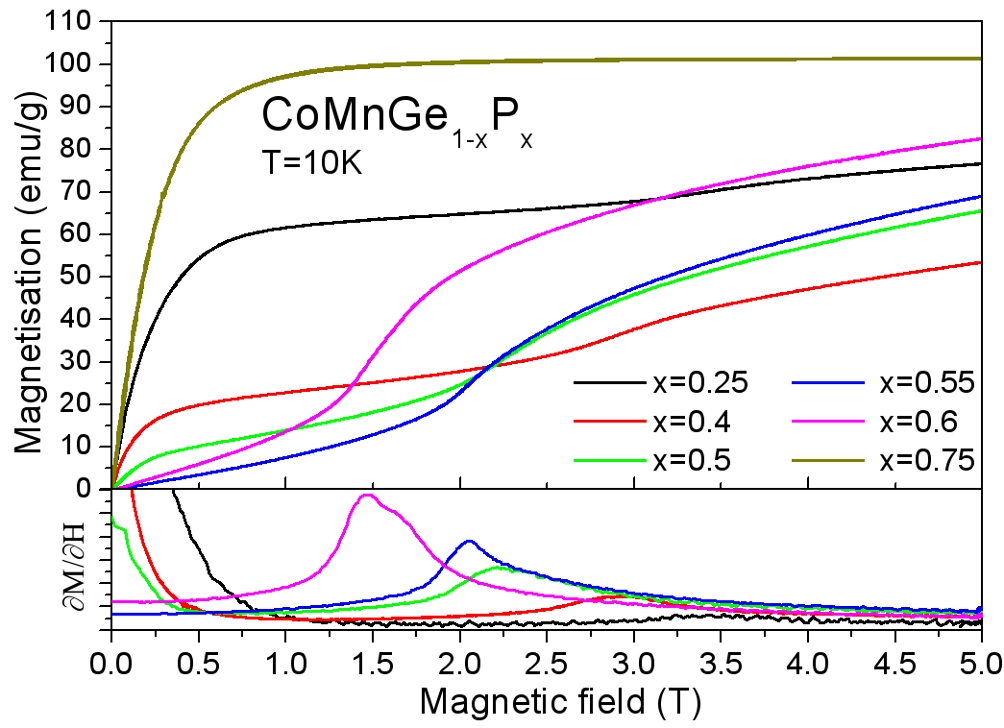} \caption{\label{fig:Mag1} (Color online) Magnetisation loops of CoMnGe$_{1-x}$P$_{x}$
at 10K.}

\end{figure}

The composition dependence of magnetisation loops taken at 10~K is
shown in Fig.~\ref{fig:Mag1}. Metamagnetism is most visible in the
single phase compositions with \emph{x}=0.55 and 0.6. In these samples,
the initial magnetisation varies almost linearly with applied field
below the inflection point that occurs at a critical field value.
Although this critical behaviour is also visible in all compositions
with $x\leq0.6$, the increasing volume of ferromagnetic CoMnGe phase
at low \emph{x} suppresses the sharpness of the upturn in magnetisation.
A distinct shift of the peak of $\frac{\partial M}{\partial H}$,
seen in Fig.~\ref{fig:Mag1}b shows that the critical field decreases
sharply as more Ge is replaced by P (towards small \emph{x}).

\begin{figure}
\includegraphics[width=85mm]{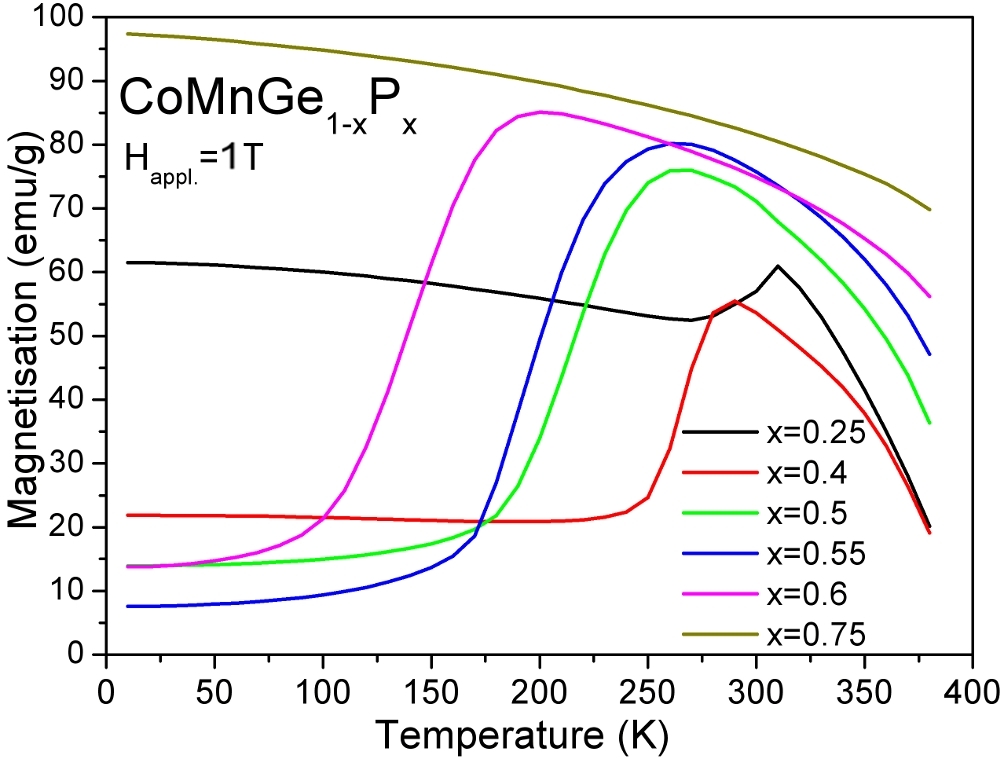}\caption{\label{fig:Mag2} (Color online) Iso-field magnetisation of CoMnGe$_{1-x}$P$_{x}$
as a function of temperature, measured in a 1~Tesla applied field.
We observe metamagnetism in almost all samples, and a sharp change
in the metamagnetic critical temperature with composition in the range
\emph{x}=0.4 to 0.6.}

\end{figure}

The composition dependence of the critical field in a field of 1~Tesla
is shown in Fig.~\ref{fig:Mag2} and is consistent with the magnetisation
loops collected at 10~K. The lower the critical field that is observed
at 10~K (Fig. \ref{fig:Mag1}), the lower the critical temperature
is at which the 1~T applied magnetic field is sufficiently large
to bring the sample to the high magnetic state. A strong dependence
of the metamagnetic transition temperature on crystal structure is
evident from this Figure, when compared with Fig.~\ref{fig:XRD_abc}.
Although samples with $x\leq0.5$ contain a minor second phase, they
all show a characteristic increase in the magnetisation as a function
of temperature, indicative of native metamagnetism. In the high Ge
(low \emph{x}) side of the series, with \emph{x}=0.25, although the
substantial ferromagnetic volume (from the CoMnGe phase) largely suppresses
the metamagnetic transition, it still appears at around room temperature
(RT). 

As the samples become richer in phosphor, the critical temperature
decreases sharply from $\sim$300K to $\thicksim$80K and it eventually
disappears for \emph{x}=0.75. The highest magnetisation values in
a 1~T applied field show an increase with \emph{x} (except for \emph{x}=0.25)
as the result of a balance of several effects. Firstly, it is easily
foreseeable that at lower transition temperatures ferromagnetic configurations
will exhibit larger overall net moments compared to those of the higher
temperature ones. Secondly, the presence of the hexagonal CoMnGe lifts
the low temperature {}``baseline'' of magnetisation in Fig. \ref{fig:Mag2}.

\section{Summary and Conclusions\label{sec:Conclusions}}

Using DFT calculations based on a {}``prototype'' binary MnP composition,
we have investigated the occurrence of AFM and FM states in Mn-based
orthorhombic (\emph{Pnma}, 62) alloys. As the result of isotropic
expansion, the FM(1) ground state is no longer stable but instead
AFM coupling of the spins on the Mn atoms is predicted above $d{}_{1}\gtrsim$2.95~Å~\citep{Gercsi_MnP}.
In this work, we have extended our theoretical investigation to higher
hydrostatic expansions and found the re-occurrence of ferromagnetism
(FM2) at large Mn-Mn separations over $d{}_{1}\gtrsim$3.37~Å that
also explains the collinear ferromagnetism in CoMnGe with $d{}_{1}=3.4$Å. 

Based solely on the deduced magnetic stability plot (Fig. \ref{fig:StabilityandM}),
we designed a series of pseudo-ternary CoMn-based alloys in order
to experimentally prove the validity of our theoretical concept. Taking
two collinear FM ternaries: one, CoMnP with a low $d{}_{1}$ from
the FM1 region ; the other, CoMnGe with a high $d{}_{1}$ from the
FM2 region we attempted to drive the alloy magnetism towards the metamagnetic/AFM
zone by careful structural design. 

The experimental investigation of CoMnGe$_{1-x}$P$_{x}$ has indeed
revealed an AFM ground state for compositions $x\thickapprox0.5$.
The appearance of a magnetic field- and temperature-dependent metamagnetic
transition in several samples also suggests the existence of complex
non-collinear spin structure in most of them , and in particular in
the range \emph{x}=0.4 to 0.6. The large predicted $N_{Tot}(E_{\mathbb{F}})$
for $x=0.5$ (in Sec. \ref{sec:Theoretical-Results}) in a hypothetical
FM state is because of a shift in the energy of the hybridisation-derived
pseudogap as the lattice parameters expand upon Ge substitution for
phosphor. 

Although the complex magnetic spin structure of these new samples
is to be determined, the system is an example of the stabilization
of non-collinear magnetism through the formation of hybridization
gap at the Fermi energy as described by \citet{Eriks_bandcross} and
as recently found in the ternary CoMnSi \citep{AlexHRPD}. The Mn-containing
\emph{Pnma} structure is extremely versatile with regard to elemental
substitution. The above demonstration of a structurally-directed tuning
of magnetic properties therefore provides a potential direction for
future tailoring of metamagnetic phase transitions towards their use
in applications such as those that rely on the magnetocaloric effect.
\begin{acknowledgments}
Z. G. is grateful for the invitation and financial support of NIMS
through the Open Research Institute Program. K.G.S. acknowledges financial
support from The Royal Society. Furthermore, the authors thank H.
S. Amin for his assistance in the SEM sample preparation and analysis.
The research leading to these results has received funding from the
European Community's 7th Framework Programme under grant agreement
No. 214864 ({}``SSEEC''). Computing resources provided by Darwin
HPC and Camgrid facilities at The University of Cambridge and the
HPC Service at Imperial College London are gratefully acknowledged.
\end{acknowledgments}
\bibliographystyle{revtex4/apsrev}
\addcontentsline{toc}{section}{\refname}\bibliography{CoMnPGe}

\end{document}